\documentclass[aps,prl,superscriptaddress,amsfonts,11pt]{revtex4-2}

\usepackage{graphicx}%
\usepackage{dcolumn}
\usepackage{amsmath}
\usepackage{color}
\usepackage{multirow}

\makeatletter
\def\btt#1{\texttt{\@backslashchar#1}}%
\DeclareRobustCommand\bblash{\btt{\@backslashchar}}%
\makeatother

\definecolor{fj_color}{cmyk}{1, 0.3, 0, 0}
\definecolor{cwh_color}{cmyk}{0, 0.8, 0.8, 0}
\definecolor{hr_color}{cmyk}{0.5, 0.8, 0.0,0}

\newcommand{\SRO}{Sr\textsubscript{2}RuO\textsubscript{4}}

\begin{document}

\preprint{PREPRINT (\today)}

\title{$\mu$SR measurements on Sr$_2$RuO$_4$ under $\langle 110 \rangle$ uniaxial stress}

\author{Vadim Grinenko}
 \email{vadim.a.grinenko@gmail.com}
 \affiliation{Institute for Solid State and Materials Physics, Technische Universit\"{a}t Dresden, D-01069 Dresden, Germany}
 \affiliation{Tsung-Dao Lee Institute, Shanghai Jiao Tong University, Pudong, Shanghai, China}

\author{Rajib Sarkar}
\affiliation{Institute for Solid State and Materials Physics, Technische Universit\"{a}t Dresden, D-01069 Dresden, Germany} 

  \author{Shreenanda Ghosh}
\affiliation{Institute for Solid State and Materials Physics, Technische Universit\"{a}t Dresden, D-01069 Dresden, Germany} 
\affiliation{Institute for Quantum Matter, William H. Miller III Department of Physics and
Astronomy, The Johns Hopkins University, Baltimore, Maryland 21218, USA}

\author{Debarchan Das}
  \affiliation{Laboratory for Muon Spin Spectroscopy, Paul Scherrer Institut, CH-5232 Villigen PSI, Switzerland}

\author{Zurab Guguchia}
  \affiliation{Laboratory for Muon Spin Spectroscopy, Paul Scherrer Institut, CH-5232 Villigen PSI, Switzerland}
  
 \author{Hubertus Luetkens}
  \affiliation{Laboratory for Muon Spin Spectroscopy, Paul Scherrer Institut, CH-5232 Villigen PSI, Switzerland}
  
  \author{Ilya Shipulin}
\affiliation{Leibniz Institute for Solid State and Materials Research, D-01069, Dresden, Germany}

  \author{Aline Ramires}
 \affiliation{Paul Scherrer Institut, CH-5232 Villigen PSI, Switzerland}

 \author{Naoki Kikugawa}
  \affiliation{National Institute for Materials Science, Tsukuba 305-0003, Japan}

\author{Yoshiteru Maeno}
  \affiliation{Department of Physics, Kyoto University, Kyoto 606-8502, Japan}
 \affiliation{Toyota Riken - Kyoto-Univ. Research Center (TRiKUC), Kyoto 606-8501, Japan}
  
 \author{Kousuke Ishida}
 \affiliation{Max Planck Institute for Chemical Physics of Solids, D-01187 Dresden, Germany}
 
 \author{Clifford W. Hicks}
   \affiliation{Max Planck Institute for Chemical Physics of Solids, D-01187 Dresden, Germany}
  \affiliation{School of Physics and Astronomy, University of Birmingham, Birmingham B15 2TT, United Kingdom}

\author{Hans-Henning Klauss}
  \affiliation{Institute for Solid State and Materials Physics, Technische Universit\"{a}t Dresden, D-01069 Dresden, Germany}

\begin{abstract}
	Muon spin rotation/relaxation ($\mu$SR) and polar Kerr effect measurements provide  evidence for a time-reversal symmetry breaking (TRSB) superconducting state in Sr$_2$RuO$_4$. However, the absence of a cusp in the superconducting transition temperature ($T_{\rm c}$) vs. stress and the absence of a resolvable specific heat anomaly at TRSB transition temperature ($T_{\rm TRSB}$) under uniaxial stress challenge a hypothesis of TRSB superconductivity. Recent $\mu$SR studies under pressure and with disorder indicate that the splitting between $T_{\rm c}$ and $T_{\rm TRSB}$ occurs only when the structural tetragonal symmetry is broken. To further test such behavior, we measured $T_\text{c}$ through susceptibility measurements, and $T_\text{TRSB}$ through $\mu$SR, under uniaxial stress applied along a $\langle 110 \rangle$ lattice direction. We have obtained preliminary evidence for suppression of $T_\text{TRSB}$ below $T_\text{c}$, at a rate much higher than the suppression rate of $T_\text{c}$. 
\end{abstract}

\maketitle


\section{Introduction}
Even after nearly 30 years of research, the superconductivity of \SRO{} is a mystery \cite{Maeno94_Nature}. 
The greatest conundrum is the evidence that the order parameter combines even parity with time-reversal symmetry breaking. 
The evidence for even parity comes especially from recent NMR measurements~\cite{Pustogow19_Nature, Ishida20_JPSJ, Chronister21_PNAS}, showing a spin-singlet-like susceptibility drop below $T_\text{c}$. 
Evidence for time-reversal symmetry breaking (TRSB) comes from anomalous switching noise in junctions~\cite{Anwar13_SciRep, Nakamura12_JPSJ}, Kerr rotation~\cite{Xia06_PRL}, and enhanced muon spin relaxation in the superconducting state~\cite{Luke98_Nature, Luke2000, Shiroka2012, Higemoto2016, Grinenko21_NatComm, Grinenko21_NatPhys, Huddart2021}. 

In previous muon spin rotation/relaxation ($\mu$SR) studies of \SRO{}, some of the present authors observed that $T_\text{TRSB}$ and $T_\text{c}$ split under uniaxial stresses applied along a $\langle 100 \rangle$ crystallographic direction, thereby supporting a scenario of TRSB superconductivity~\cite{Grinenko21_NatPhys}. Together with the observation that $T_\text{TRSB}$ and $T_\text{c}$ track each other under hydrostatic stress and also under an introduction of disorder~\cite{Grinenko21_NatComm}, this observation suggests a symmetry-protected even-parity chiral superconducting state, $d_{xz} \pm id_{yz}$. This order parameter would be surprising because the line node at $k_z=0$ implies, conventionally, interlayer pairing, while the interlayer interactions in \SRO{} are expected to be weak given the apparent strong two-dimensionality of its electronic structure \cite{Bergemann03_AIP}. 
Recently, it has been proposed that the superconductivity of \SRO{} could emerge through interorbital interactions~\cite{Suh20_PRR, Clepkens21_PRR, Gingras19_PRL, Kaeser21_arxiv, Beck21_arxiv, Romer2022}. Pairing in these proposals is primarily driven by local interactions (such as Hund’s coupling), evading the need for strong inter-layer coupling and suggesting that chiral $d$-wave superconductivity could be a rather natural order parameter for \SRO{}.

 So far, the splitting of the transitions under uniaxial $\langle 100 \rangle$ stress has been seen only in  $\mu$SR measurements. In contrast with expectations for a chiral state, a second anomaly was not resolved either in heat capacity~\cite{Li21_PNAS} nor in elastocaloric effect~\cite{Li2022} measurements in which uniaxial stress was applied to split $T_\text{c}$ and $T_\text{TRSB}$. In addition, the expected cusp in the dependence of $T_\text{c}$ on uniaxial stress has not been resolved for either $\langle 100 \rangle$ or $\langle 110 \rangle$ directions. These contradictory results have led to proposals that TRSB in \SRO{} is finely-tuned with order parameters of the form $s \pm id$, $s \pm ip$  \cite{Romer19_PRL, Romer21_arXiv} or $d \pm ig$ \cite{Kivelson20_npj, Wagner20_PRB}, or even might occur only in the vicinity of extended defects~\cite{Willa21_PRB}.

For $\langle 100 \rangle$  stress, $T_\text{c}$ strongly increases on approach to a van Hove singularity \cite{Hicks14_Science}, while $T_\text{TRSB}$ barely changes \cite{Grinenko21_NatPhys}. The high sensitivity of the electronic structure to $\langle 100 \rangle$ stress complicates the interpretation of the observed splitting of the transitions. In contrast, the electronic band structure is less sensitive to stress applied along  $\langle 110 \rangle$ axes. Although a weaker dependence on stress is expected, it may be easier to interpret observed effects. 

Under the tetragonal lattice symmetry of \SRO{}, 
$T_\text{c}$ and $T_\text{TRSB}$ are expected to split under shear strain $\varepsilon_6$, which has $\langle 110 \rangle$ principal axes, for a $d_{xz} \pm id_{yz}$ order parameter.  In the limit of small strain, the rates of change $|dT_\text{c}/d\varepsilon_6|$ and $|dT_\text{TRSB}/d\varepsilon_6|$ are inversely proportional to the condensation energies associated with each phase transition \cite{Grinenko21_NatPhys}. Here, we report measurements of both $T_\text{c}$ and $T_\text{TRSB}$ under uniaxial stress applied along $\langle 110 \rangle$ lattice direction. $T_\text{c}$ was measured through magnetic susceptibility, and $T_\text{TRSB}$ through $\mu$SR. $\langle 110 \rangle$ stress, in addition applying shear strain, also affects the unit cell volume and lattice constant ratio $c/a$. The effect of these strain components on $T_\text{c}$ and $T_\text{TRSB}$ was estimated using the elasticity stiffness matrix known from the ultrasound experiments \cite{Ghosh21_NatPhys} and experimental dependencies of $T_\text{c}$ on hydrostatic pressure and uniaxial $c$-axis strain.  
We obtained preliminary evidence that $T_\text{TRSB}$ and $T_\text{c}$ split under the $\langle 110 \rangle$ uniaxial stress, with the condensation energy associated with the time-reversal symmetry breaking being very small compared to that associated with the superconductivity overall. These are very challenging measurements due to the small size of the signal. We publish a preliminary data set now, because a more authoritative data set might not be possible for some time, mainly due to limited beamtime. 

\section{Experimental design and Results}

Single crystals of Sr$_2$RuO$_4$ were grown by a floating zone method~\cite{Bobowski19_CondMat}. Data from two samples, labeled A and B, are reported. 
In order to obtain samples of sufficient length for the uniaxial stress apparatus, samples were cut from a rod that grew nearly along a $\langle 110 \rangle$ lattice direction. The samples studied here were either cleaved or ground into plates, exposing the interior of the as-grown rod to the muon beam.

Samples were mounted into holders using Stycast 2850 epoxy as shown in Ref.~\cite{Grinenko21_NatPhys}. The epoxy layers were generally 50--100~$\mu$m thick. Additional steps were taken to improve the chances of reaching high stresses without fracturing the sample.  (1) They were cut at a $10^\circ$ angle with respect to the $ab$ plane, so that
shear stresses in the sample do not align with cleave planes. (2) 10~$\mu$m-thick titanium foils were affixed to their surfaces with Stycast 1266 for sample B for mechanical reinforcement. (3) The slots in the holder were chamfered, as shown in Fig.~1(b) of Ref.~\cite{Grinenko21_NatPhys}, to smooth the interface between the free and clamped portions of the sample. For further details see Refs.~\cite{Grinenko21_NatPhys,Ghosh2020,ghosh2021manipulation}.

The specific heat of a small piece of the crystal remaining after the cutting of the plates for $\mu$SR sample A was measured using the thermal relaxation method in a physical property measurement system (PPMS, Quantum Design). The ac susceptibility of the samples prepared for the $\mu$SR experiments was measured in situ using pairs of concentric coils located behind the samples. The coils were wound on each other, one of which was used as excitation and the other as a pick-up coil. The applied field for the susceptibility measurements was $\sim 10$~$\mu$T. For further details, see Ref.~\cite{Grinenko21_NatPhys,ghosh2021manipulation}.

\begin{figure}[tbh]
	\centering
	\includegraphics[width=1\linewidth]{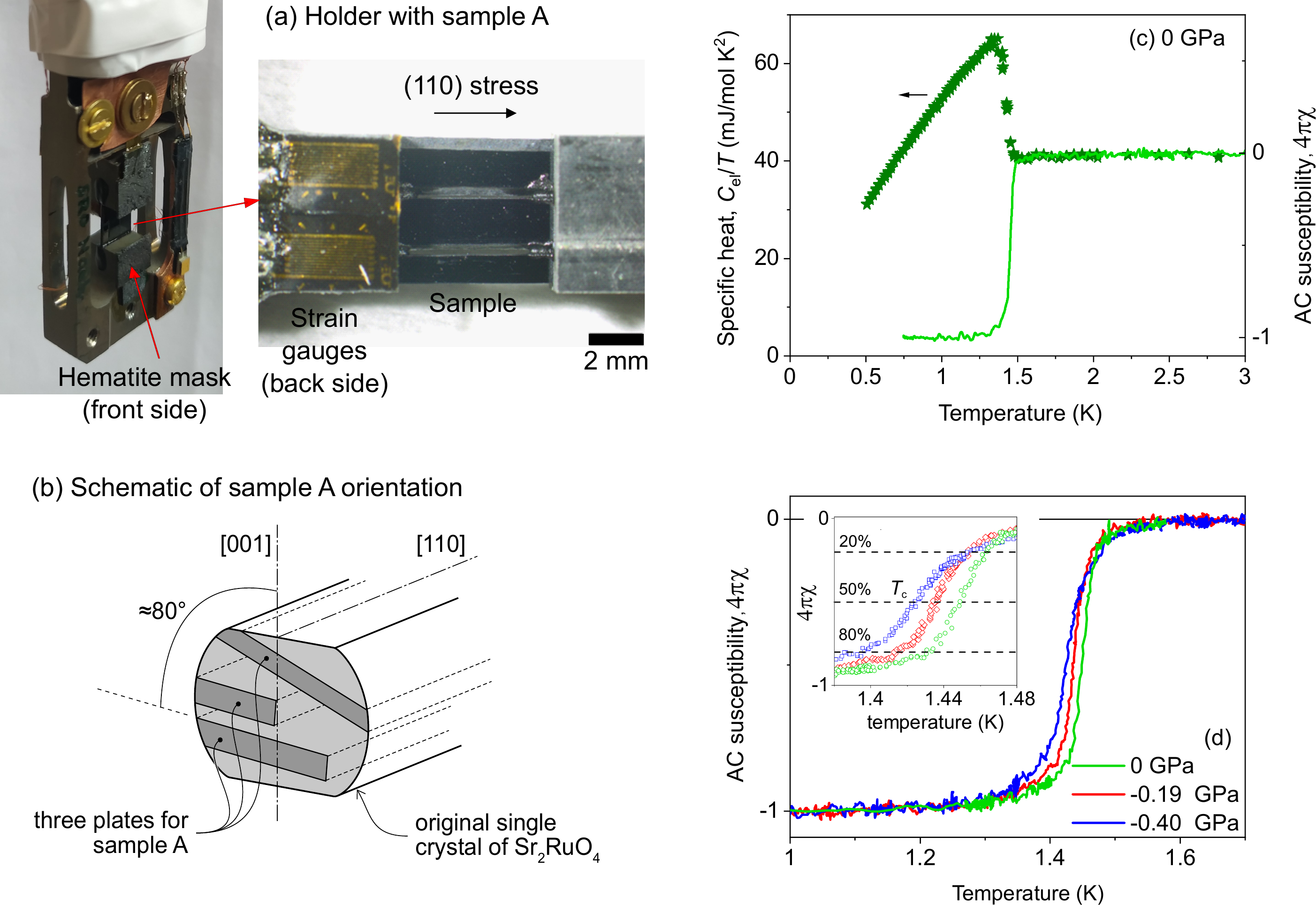}
	\caption{{\bf (a)} Photographs of the holder with a mounted sample A consisting of three rectangular-shaped pieces glued into the holder side by side. (left) The front view of the mounted sample with hematite masks used to screen the portions of the holder in the muon beam. Concentric coils behind the sample (not seen) are used for in situ measurements of $T_\text{c}$. (right) The photograph of the sample from the back side with two glued strain gauges.  {\bf (b)} Schematics of the Sr$_2$RuO$_4$ rod crossection illustrating the orientation of the individual pieces, determined by X-ray Laue photos. The misalignment of the pieces regarding the crystallographic $c$-axis is intended to reduce the probability of a cleavage under stress. {\bf (c)} Comparison of the temperature dependence of the specific heat and in situ ac susceptibility measured at zero stress. {\bf (d)}  The temperature dependencies of the in situ ac susceptibilities at different applied uniaxial stress. Inset shows zoom a region close to $T_{\rm c} = 1.448 \pm 0.015$ K, which we take as the transition midpoint. Heat capacity and transverse-field $\mu$SR data show that the samples are fully superconducting, so we identify the extrema of
		the susceptibility signal as $4\pi\chi = 0$ and $-1$.}
	\label{fig:Sample-holder}
\end{figure}

The essential experimental setup was the same as those described in Refs.~\cite{Grinenko21_NatPhys, Ghosh2020, ghosh2021manipulation}. The samples are plates thick enough to stop the muon beam, mounted in a holder that facilitates application of force. A photograph of sample A, mounted in the holder, is shown in Fig.~\ref{fig:Sample-holder}a. Sample A consists of three plates cut from one single crystal, as shown in panel (b). In panel (c) the temperature dependence of the specific heat measured on a small cut-off is compared with the in situ susceptibility measurements. There is a good agreement between both sets of measurements with the zero stress $T_{\rm c} = 1.448 \pm 0.015$ K of sample A, defined as shown in panel (d).  We also performed $\mu$SR measurements on sample B, which had similar $T_{\rm c} = 1.46 \pm 0.09$ K, with a slightly broader transition width (Fig.~\ref{fig3}a). The high $T_{\rm c}$, comparable to that of clean-limit \SRO{}~\cite{Kikugawa04_PRB_1}, and the sharp superconducting transition indicate a high sample quality. The stress values were measured by a set of strain gauges mounted on the uniaxial pressure cell as described in Refs.~\cite{Grinenko21_NatPhys, Ghosh2020, ghosh2021manipulation}.

\begin{figure}
	\centering
	\includegraphics[width=0.7\linewidth]{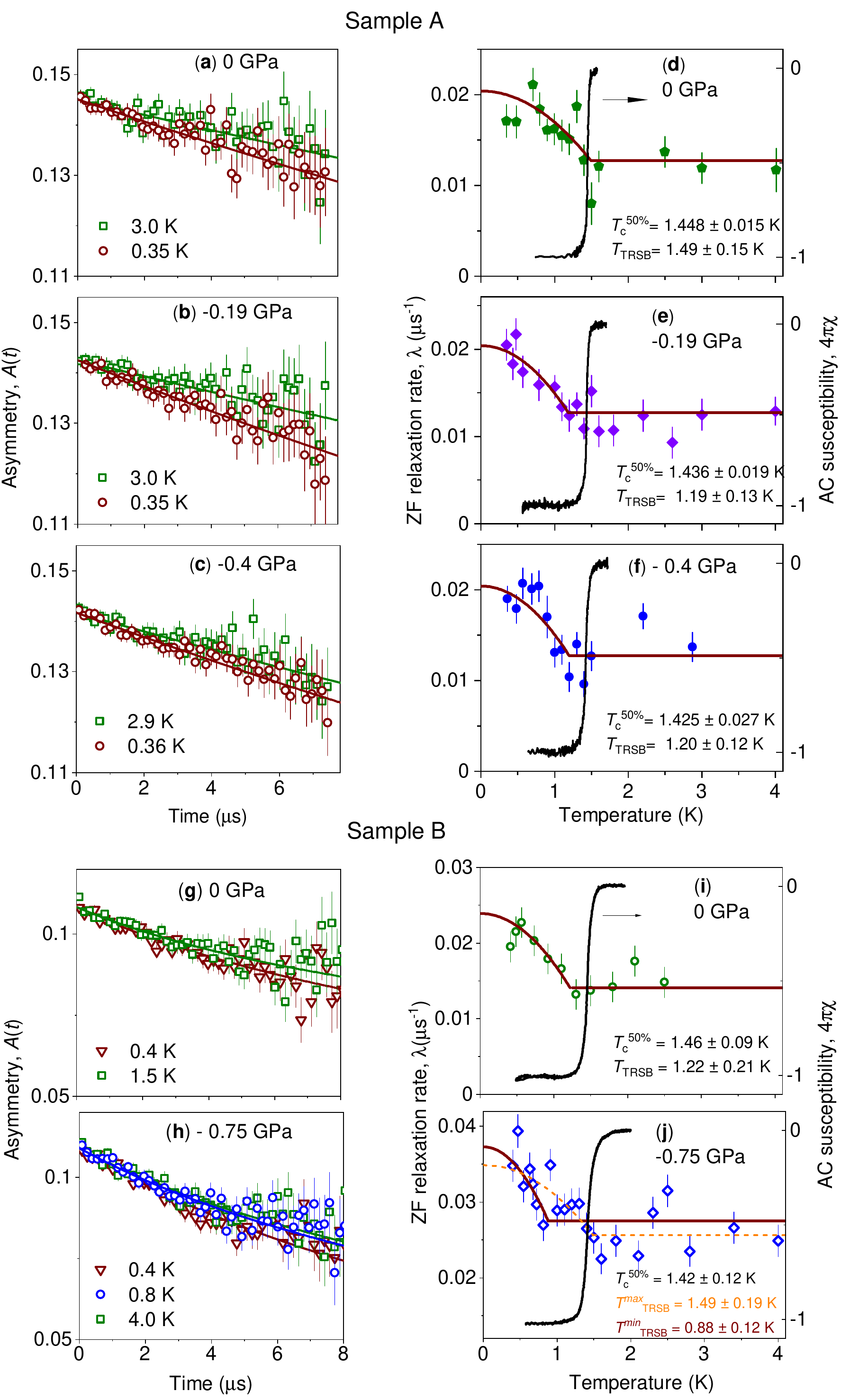}
	\caption{\textbf{Sample A - (a--c)}: Zero-field $\mu$SR asymmetry $A(t)$ at a temperature above
		$T_\text{c}$, and at the lowest temperature reached at 0~GPa,  $-0.19$~GPa and  $-0.4$~GPa $\langle 110 \rangle$ compressive stress.	\textbf{(d--f)}: Temperature dependence of the muon spin relaxation rate $\lambda$ (left), and magnetic susceptibility (right). The fits to $\lambda(T)$ (maroon lines) are explained in the text.  \textbf{Sample B -}: Zero-field $\mu$SR asymmetry $A(t)$ at a temperature above and below $T_\text{c}$ \textbf{(g)} zero stress and \textbf{(h)} -0.75 GPa $\langle 110 \rangle$ compressive stress. \textbf{(i, j)}: Temperature dependence of the muon spin relaxation rate $\lambda$ (left), and in-situ diamagnetic susceptibility data (right). Maroon solid lines are the fits to $\lambda(T)$ with $b$ as a common fitting parameter for panels (i, j); dashed orange curve in panel (j) is the fit with $b$ as an individual fitting parameter. For further details, see the text.}
	\label{fig2}
\end{figure}

\begin{figure}
	\centering
	\includegraphics[width=1\linewidth]{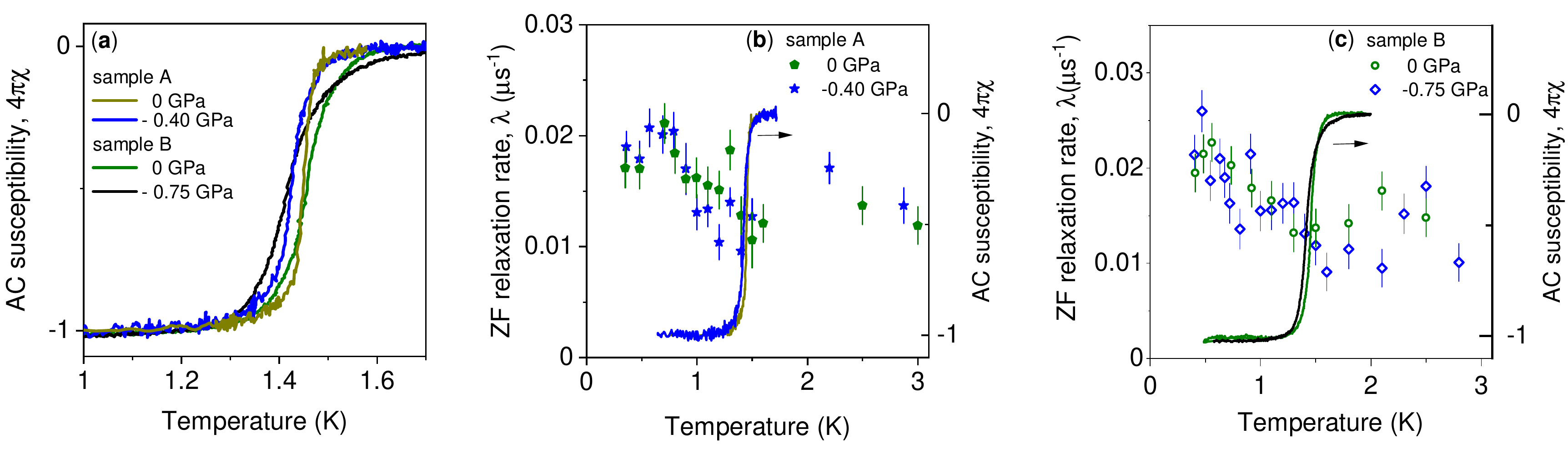}
	\caption{ {\bf (a)} Comparison of the temperature dependence of the in situ ac susceptibility between samples A and B. \textbf{(b, c)} Temperature dependencies of the muon spin relaxation rate at 0 GPa and -0.40 GPa for sample A, and at 0 GPa and -0.75 GPa for sample B, plotted together. Note that the -0.75 GPa curve for sample B was shifted down for better comparison with 0 GPa data.}
	\label{fig3}
\end{figure}

In the $\mu$SR method, spin-polarised muons are implanted into the sample, where each then precesses in its local field. The measured quantity is the decay positron emission asymmetry $A(t)$, which is proportional to the muon spin polarisation at time $t$. In this work, we performed experiments in zero external magnetic field, which was dynamically compensated to fields smaller than 1~$\mu$T. Asymmetry curves $A(t)$ at two temperatures, one above and one well below $T_{\rm c}$,  are shown in Fig.~\ref{fig2}, panels (a--c) for zero force and two applied pressures for sample A and in panels (g, h) for sample B. An increase in the muon spin relaxation rate at low temperatures is observed for all applied pressures, indicating the presence of time-reversal symmetry breaking as seen in previous studies \cite{Luke98_Nature,Luke2000,Shiroka2012, Higemoto2016,Grinenko21_NatPhys,Grinenko21_NatComm}.   
Following the previously established procedure in Ref.~\cite{Grinenko21_NatPhys}, 
the exponential muon spin relaxation rate $\lambda$ at each temperature is obtained by fitting:
\begin{equation}
	A_\text{fit}(T, t) = A_\text{sam} e^{-\lambda(T) t} + A_\text{bkg}.
\end{equation}
$A_\text{bkg}$ is a background constant to account for muons that implant into non-superconducting material such as cryostat walls, and $A_\text{sam}$ is the sample signal strength. $A_\text{bkg}$ and $A_\text{sam}$ are determined from weak transverse-field $\mu$SR measurements as described in Ref.~\cite{Grinenko21_NatPhys} Thus, in the analysis of ZF data $\lambda$ is the sole free fitting parameter.
Previously, we studied the background relaxation rate by measuring the holder without a sample~\cite{Grinenko21_NatPhys}. We found that within the error bars of the measurements, $A_\text{bkg}$ is temperature-independent. However, the absolute value of the $A_\text{bkg}$ is sensitive to the holder and Hematite mask position.

Results for sample A at zero stress and two compressive stresses are shown in Fig.~\ref{fig2}, panels (d--f).  
It is seen that at each stress the muon spin relaxation rate ($\lambda$) is enhanced in the superconducting state. To extract $T_\text{TRSB}$, we fit the temperature dependence of $\lambda(T)$ at each stress using a phenomenological equation:
\begin{equation}
	\lambda(T) = \left\{
	\begin{array}{ll}
		\lambda_0 + b \times ( 1 - (T/T_\text{TRSB})^2), & T < T_\text{TRSB} \\
		\lambda_0, & T > T_\text{TRSB} \\
	\end{array}
	\right.
\end{equation}
For sample A, the $b$ and $\lambda_0$ are taken to be common fitting parameters among all three stresses, while $T_\text{TRSB}$ is obtained independently at each stress. This assumption is justified because it is the same sample at each stress and the change in $T_\text{TRSB}$ is small. This fit gives $T_\text{TRSB} = 1.49 \pm 0.15$~K
at 0~GPa, $1.19 \pm 0.13$~K at $-0.19$~GPa, and $1.20 \pm 0.12$~K at $-0.4$~GPa. To allow more direct comparison, we also show the data measured at 0~GPa and $-0.4$~GPa in one plot without fitted curves in Fig.~\ref{fig3}(b).

The analysis of the data for sample B, shown in Fig.~\ref{fig2} panels (i, j), with the same model results in $T_\text{TRSB} = 1.22 \pm 0.21$~K at 0~GPa and $0.88 \pm 0.12$~K at $-0.75$~GPa. The rapid suppression of $T_\text{TRSB}$ is consistent with the result from sample A. 
However, the uncertainties must be considered. For an unknown reason (possibly a shift in the sample and/or the hematite mask position), $\lambda_0$ changed substantially when stress was applied, so we had to make $\lambda_0$ a stress-dependent fitting parameter. The analysis performed in Ref.~\cite{Grinenko21_NatPhys} indicates that $\lambda_0$ is sensitive to the precise sample configuration, which in general, might be modified by the application of force, while $b$, characterizing the strength of spontaneous magnetic fields, is likely set by the defect density – it varies from sample to sample (see Ref.~\cite{Grinenko21_NatComm} for discussion). So the defect density is unlikely to change with stress. (At first, the measurements under stress were performed. Then the stress was slowly released at 6K, and the zero-stressed measurements were performed.) For a TRSB superconductor, the magnitude of $b$ can also depend on the size of the superconducting gaps, and hence it might depend on $T_\text{TRSB}$ and $T_\text{c}$ values. If $b$ is taken to be stress-dependent, we obtain a larger $T_\text{TRSB}$ at $-0.75$~GPa than at $0$~GPa, as shown in panel (j) by the dashed orange curve giving a very high $T^{\rm max}_\text{TRSB} = 1.49 \pm 0.12$~K. This would result in an opposite $T_\text{TRSB}$ dependence on stress for sample B compared to sample A, which seems unlikely. To allow more direct comparison, we also show the data for sample B measured at 0~GPa and $-0.75$~GPa in one plot without fitted curves in Fig.~\ref{fig3}c. The -0.75 GPa curve for sample B was shifted down for better comparison with 0 GPa data.

\begin{figure}
	\centering
	\includegraphics[width=0.7\linewidth]{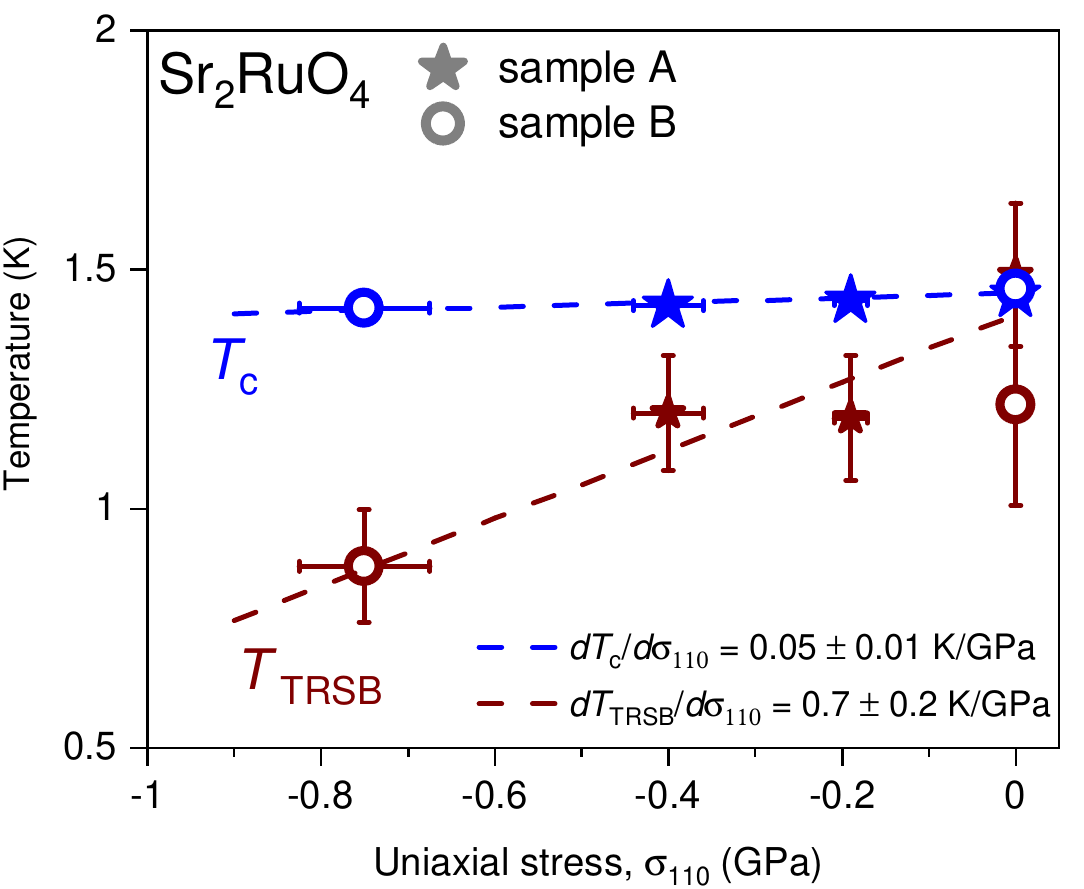}
	\caption{The stress dependence of the superconducting and TRSB transition temperatures extracted from the experimental data together with linear fitting curves to extract $|{dT_\text{\rm c}}/{d\sigma_{110}}|$ and the upper limit for the $|{dT_\text{\rm TRSB}}/{d\sigma_{110}}|$ slope.}
	\label{fig4}
\end{figure}

The resulting $T-\sigma_{110}$ experimental phase diagram is shown in Fig.~\ref{fig4}. Since there is the discussed uncertainty in the analysis for sample B, the dashed maroon curve represents the upper limit for the $|dT_\text{TRSB}/d\sigma_{110}|$ slope, which substantially exceeds $|dT_\text{c}/d\sigma_{110}|$. However, with the current data, we cannot exclude the possibility that $T_\text{TRSB}$ and $T_\text{c}$ do not split under $\langle 110 \rangle$ stress. 

\section{Discussion}

Our measurements show $\frac{dT_{\rm c}}{d\sigma_{110}} = 0.05 \pm 0.01$~K/GPa , $T_{\rm c}$ decreases slightly under the $\langle 110 \rangle$ compression (Fig.~\ref{fig4}). However, uniaxial stress applied along the $\langle 110 \rangle$ direction induces not only shear strain, but also a change in the unit cell volume and in the lattice parameter ratio $c/a$. In an analysis given in the Appendix, these latter two effects are shown to contribute about 0.1 K/GPa. This sets an upper limit on the coupling of $T_{\rm c}$ to shear strain of $|\frac{dT_{\rm c}}{d\epsilon_{6}}| \lesssim $ 5 K.

Our experimental data show that $T_{\rm TRSB}$ can be more susceptible to the stress than $T_{\rm c}$ giving an upper limit for the slope  $\frac{dT_{\rm TRSB}}{d\sigma_{110}}\approx 0.7$~K/GPa (see Fig.~\ref{fig4}). However, we cannot rule out the possibility that there is in fact no the $\langle 110 \rangle$ stress induced splitting between $T_{\rm c}$ and $T_{\rm TRSB}$. 0.7 K/GPa is much larger than the contribution from non-shear elements of the applied strain, and so would require substantial coupling to the shear strain. It is shown in the Appendix that this corresponds to an upper limit $\frac{dT_{\rm TRSB}}{d\epsilon_{6}}\approx \frac{1}{2}131 {\rm GPa}\left( \frac{dT_{\rm c}}{d\sigma_{110}}-0.1{\rm K/GPa}\right)  \lesssim $ 40 K.

Recently, two independent ultrasound studies of Sr$_2$RuO$_4$ reported a discontinuity in the shear elastic modulus $c_{\rm 66}$ at $T_{\rm c}$ in accord with expectations for a multicomponent component order parameter \cite{Ghosh21_NatPhys,Benhabib20_NatPhys}. If the order parameter is chiral, with two components related by symmetry ($p_x \pm ip_y$ or $d_{xz} \pm id_{yz}$), an Ehrenfest relationship applies:
\begin{equation}
	\Delta c_{66} = - \frac{\Delta C}{T_\text{c}} \left| \frac{dT_\text{c}}{d\varepsilon_6}\right| \left| \frac{dT_\text{TRSB}}{d\varepsilon_6}\right|,
\end{equation}
where $\Delta c_{66}$ is the change in $c_{66}$ at $T_\text{c}$, and $\Delta C$ is the heat capacity jump at $T_\text{c}$ \cite{Ghosh21_NatPhys}. 
We note that the ultrasound experiments are consistent with two-component order parameters that either break or do not break time-reversal symmetry. However, in light of other evidence reviewed in the introduction, we assume that the jump in $c_{66}$ is associated with time-reversal symmetry breaking.

In Ghosh \textit{et al} and Benhabib \textit{et al}, $\Delta c_{66} = 1.05$ and 0.03~MPa were observed, respectively~\cite{Ghosh21_NatPhys, Benhabib20_NatPhys}. 
(In Benhabib \textit{et al}, a change in shear sound velocity of 0.2~ppm was reported, indicating a change in $c_{66}$ of 0.4~ppm.) 
The reason for the difference is not clear, although the very different frequencies of the measurements (2 versus 169~MHz) are a possible factor. Taking $\Delta C = 36$~mJ/mol-K~\cite{Ghosh21_NatPhys} in Eq.~3, $\Delta c_{66} = 1.05$~MPa implies a product of slopes $|\frac{dT_\text{c}}{d\varepsilon_6}| \times |\frac{dT_\text{TRSB}}{d\varepsilon_6}| = 2500$~K$^2$, and $\Delta c_{66} =  0.03$~MPa implies a product of slopes $|\frac{dT_\text{c}}{d\varepsilon_6}| \times |\frac{dT_\text{TRSB}}{d\varepsilon_6}| = 70$~K$^2$. 
Our data provides an upper limit on $|\frac{dT_\text{c}}{d\varepsilon_6}| \times |\frac{dT_\text{TRSB}}{d\varepsilon_6}|$ of 200~K$^2$. According to Eq.~3, the value is reconcilable with $\Delta c_{66} = 0.03$~MPa \cite{Benhabib20_NatPhys}, but not $\Delta c_{66} = 1.05$~MPa \cite{Ghosh21_NatPhys}.  

Our data indicate that $T_{\rm TRSB}$ might be  considerably more sensitive to $\langle 110 \rangle$ stress than $T_{\rm c}$. If the order parameter is symmetry-protected chiral, then the ratio of heat capacity anomalies in the mean field is inverse to the ratio of slopes $|{dT_\text{c}}/{d\varepsilon_6}|$ and $|{dT_\text{TRSB}}/{d\varepsilon_6}|$ (see supplementary information in Ref.~\cite{Grinenko21_NatPhys}). If the order parameter is one with accidental degeneracy then this exact relation no longer holds, but an approximately inverse relationship between condensation energy and slopes is still expected. Our data indicate that $|dT_\text{TRSB}/d\varepsilon_6|$ could be an order of magnitude larger than $|dT_\text{c}/d\varepsilon_6|$, in which case the second heat capacity anomaly could be below the resolution limit set in Ref.~\cite{Li21_PNAS}. 

Several factors affect the ratio of slopes $|dT_\text{c}/{d\varepsilon_6}|$ and $|dT_\text{TRSB}/{d\varepsilon_6}|$. For chiral states such as $d_{xz} \pm id_{yz}$ the slopes depend on the anisotropy of the Fermi surface taking part in superconductivity. In the case of multiorbital superconductivity of \SRO~with complex Fermi surface, the calculation of the slopes is rather challenging and requires further studies.  Qualitatively, a possible explanation for a large $|{dT_\text{TRSB}}/{d\varepsilon_6}|$ and the related small heat capacity anomaly at $T_\text{TRSB}$ is that breaking time-reversal symmetry causes only a narrow node to be filled in. Recent calculations of the specific heat for various accidentally degenerate superconducting orders indicate that a better agreement between experimental data can be obtained for $s' \pm id_{\rm xy}$, with $s'$ -indicates an $s$-wave state with accidental nodes, compared to other orders \cite{Roising2022}.

In conclusion, our data set an upper limit on the dependence of $T_{\rm TRSB}$ on shear strain $\varepsilon_6$. This upper limit is compatible with $\Delta c_{66} = 0.03$~GPa, as reported in Ref.~\cite{Benhabib20_NatPhys}, but not with $\Delta c_{66} = 1.05$~GPa reported in Ref.~\cite{Ghosh21_NatPhys}. 
Further studies under a $\langle 110 \rangle$ uniaxial stress using $\mu$SR and other experimental techniques are needed to refine the obtained results.

\section{Appendix}
\renewcommand{\theequation}{A\arabic{equation}}
\renewcommand{\thefigure}{A\arabic{figure}}
\renewcommand{\thetable}{A\arabic{table}}
\setcounter{equation}{0}
\setcounter{figure}{0}
\setcounter{table}{0}

\subsection{Analysis of the applied strain}

In the experiment, we apply $\langle 110 \rangle$  stress, which results in orthorhombic distortion of the lattice and affects the unit cell parameters. In general, the applied stresses ($\sigma_{\rm i}$) coupled to strains ($\epsilon_{\rm i}$) by the elasticity stiffness matrix for tetragonal crystal symmetry:
\[
\begin{bmatrix}
	\sigma_{\rm x}  \\
	\sigma_{\rm y} \\
	\sigma_{\rm z}  \\
	\sigma_{\rm xz}  \\
	\sigma_{\rm yz}   \\
	\sigma_{\rm xy}  \\
\end{bmatrix}
=
\begin{bmatrix}
	c_{11}       & c_{12} & c_{13} & 0 & 0  & 0 \\
	c_{21}       & c_{22} & c_{23} &  0 & 0  & 0 \\
	c_{31}       & c_{32} & c_{33} &  0 & 0  & 0 \\
	\dot       & \dot & \dot &  c_{44}  & 0  & 0 \\
	\dot       & \dot & \dot &  \dot  & c_{55}  & 0 \\
	\dot       & \dot & \dot &  \dot  &  \dot & c_{66}\\
	
\end{bmatrix}
\cdot
\begin{bmatrix}
	\epsilon_{\rm x}  \\
	\epsilon_{\rm y} \\
	\epsilon_{\rm z}  \\
	2\epsilon_{\rm xz}  \\
	2\epsilon_{\rm yz}  \\
	2\epsilon_{\rm xy}  \\
\end{bmatrix},
\]   
where i = x, y, z, xz, zy, and xy. The shear strain in the main text is connected to the xy strain by $\epsilon_{6} = 2\epsilon_{\rm xy}$. Using the measured elastic constants from Ref.~\cite{Ghosh21_NatPhys}, we get:
\[
\begin{bmatrix}
	\sigma_{\rm x}  \\
	\sigma_{\rm y} \\
	\sigma_{\rm z}  \\
	\sigma_{\rm xz}  \\
	\sigma_{\rm yz}   \\
	\sigma_{\rm xy}  \\
\end{bmatrix}
=
\begin{bmatrix}
	243.9       & 137.7 & 85.0 & 0 & 0  & 0 \\
	137.7       & 243.9  & 85.0 &  0 & 0  & 0 \\
	85.0       & 85.0 & 257.2 &  0 & 0  & 0 \\
	\dot       & \dot & \dot &  69.5  & 0  & 0 \\
	\dot       & \dot & \dot &  \dot  & 69.5  & 0 \\
	\dot       & \dot & \dot &  \dot  &  \dot & 65.5\\
	
\end{bmatrix}
\cdot
\begin{bmatrix}
	\epsilon_{\rm x}  \\
	\epsilon_{\rm y} \\
	\epsilon_{\rm z}  \\
	2\epsilon_{\rm xz}  \\
	2\epsilon_{\rm yz}  \\
	2\epsilon_{\rm xy}  \\
\end{bmatrix},
\]  
where the values are given in GPa. Inverting the matrix we get the relationship between strains and stresses:

\[
\begin{bmatrix}
	\epsilon_{\rm x}  \\
	\epsilon_{\rm y} \\
	\epsilon_{\rm z}  \\
	2\epsilon_{\rm xz}  \\
	2\epsilon_{\rm yz}  \\
	2\epsilon_{\rm xy}  \\
\end{bmatrix}
=
\begin{bmatrix}
	0.00624       & -0.00317 & -0.00102 & 0 & 0  & 0 \\
	-0.00317      & 0.00624  & -0.00102 &  0 & 0  & 0 \\
	-0.00102       & -0.00102 & 0.00456 &  0 & 0  & 0 \\
	\dot       & \dot & \dot &  0.01439 & 0  & 0 \\
	\dot       & \dot & \dot &  \dot  & 0.01439  & 0 \\
	\dot       & \dot & \dot &  \dot  &  \dot & 0.01527\\
	
\end{bmatrix}
\cdot
\begin{bmatrix}
	\sigma_{\rm x}  \\
	\sigma_{\rm y} \\
	\sigma_{\rm z}  \\
	\sigma_{\rm xz}  \\
	\sigma_{\rm yz}   \\
	\sigma_{\rm xy}  \\
\end{bmatrix}.
\]
The experimental value for the slope of $T_{\rm c}$ under hydrostatic pressure is $dT_{\rm c}/d\sigma_{\rm hydro} = 0.23 \pm 0.01$ K/GPa and under uniaxial $c$-axis strain $dT_{\rm c}/d\sigma_z = 0.076 \pm 0.006$ K/GPa, where $\sigma_z$ denotes uniaxial stress along the $c$-axis, $\sigma_{\rm hydro} \equiv \sigma_{\rm x} = \sigma_{\rm y} = \sigma_{\rm z} $ hydrostatic stress, where other components are zero.

\subsection{Derivation of $\frac{dT_{\rm c}}{d\epsilon_{\rm x}}$, $\frac{dT_{\rm c}}{d\epsilon_{\rm y}}$ and $\frac{dT_{\rm c}}{d\epsilon_{\rm z}}$ }

To estimate the effect of the  $\langle 110 \rangle$ stress on $T_{\rm c}$ in the limit of small deformations one can decompose the effect of hydrostatic pressure and $c$-axis stress into individual components. To obtain the derivatives under hydrostatic pressure and $c$-axis stress we have used the stain-stress matrix:
\begin{equation}
	\frac{dT_{\rm c}}{d\sigma_{\rm hydro}} = 2\frac{dT_{\rm c}}{d\epsilon_{\rm x(y)} }\frac{d\epsilon_{\rm x(y)}}{d\sigma_{\rm hydro}}+\frac{dT_{\rm c}}{d\epsilon_{\rm z} }\frac{d\epsilon_{\rm z}}{d\sigma_{\rm hydro}} = 0.0041\frac{dT_{\rm c}}{d\epsilon_{\rm x(y)}}+0.00252\frac{dT_{\rm c}}{d\epsilon_{\rm z}},
\end{equation}
\begin{equation}
	\frac{dT_{\rm c}}{d\sigma_{\rm z}} = 2\frac{dT_{\rm c}}{d\epsilon_{\rm x(y)} }\frac{d\epsilon_{\rm x(y)}}{d\sigma_{\rm z}}+\frac{dT_{\rm c}}{d\epsilon_{\rm z} }\frac{d\epsilon_{\rm z}}{d\sigma_{\rm z}} =-0.00204\frac{dT_{\rm c}}{d\epsilon_{\rm x(y)}}+0.00456\frac{dT_{\rm c}}{d\epsilon_{\rm z}}.
\end{equation} 
Inverting the equations we get:
\begin{equation}
	\frac{dT_{\rm c}}{d\epsilon_{\rm x(y)}}  = 191.3\frac{dT_{\rm c}}{d\sigma_{\rm hydro}}-105.72\frac{dT_{\rm c}}{d\sigma_{\rm z}} \approx 35.96 {~\rm K},
	\label{Eq_epsilon_xx}
\end{equation}
\begin{equation}
	\frac{dT_{\rm c}}{d\epsilon_{\rm z}}  = 85.58\frac{dT_{\rm c}}{d\sigma_{\rm hydro}}+172.0\frac{dT_{\rm c}}{d\sigma_{\rm z}} \approx 32.76 {~\rm K}.
	\label{Eq_epsilon_zz}
\end{equation}

\subsection{Effect of the volume change and tetragonal distortions}
The effect of the hydrostatic pressure and the $c$-axis stress on $T_{\rm c}$ can also be decomposed on the fractional volume change of the unit cell $\epsilon_{\rm v} = \Delta V /V = \epsilon_{\rm x} + \epsilon_{\rm y} + \epsilon_{\rm z}$, and the fraction of the a volume-preserving tetragonal distortion $\epsilon_{\rm tet}=\epsilon_{\rm z} - (\epsilon_{\rm x} + \epsilon_{\rm y})/2$.

\begin{equation}
	\frac{dT_{\rm c}}{d\sigma_{\rm hydro}} = \frac{dT_{\rm c}}{d\epsilon_{\rm v}}\frac{d\epsilon_{\rm v}}{d\sigma_{\rm hydro}} = \frac{dT_{\rm c}}{d\epsilon_{\rm v}}(\frac{d\epsilon_{\rm x}}{d\sigma_{\rm hydro}}+\frac{d\epsilon_{\rm y}}{d\sigma_{\rm hydro}}+\frac{d\epsilon_{\rm z}}{d\sigma_{\rm hydro}})  = 0.00662\frac{dT_{\rm c}}{d\epsilon_{\rm v}}.
	\label{ewV}
\end{equation}
Thus, for the effect of the volume change we have:	$\frac{dT_{\rm c}}{d\epsilon_{\rm v}} = 34.74$ K.

\begin{equation}
	\frac{dT_{\rm c}}{d\sigma_{\rm z}} = \frac{dT_{\rm c}}{d\epsilon_{\rm v}}\frac{d\epsilon_{\rm v}}{d\sigma_{\rm z}} + \frac{dT_{\rm c}}{d\epsilon_{\rm tet}}\frac{d\epsilon_{\rm tet}}{d\sigma_{\rm z}}= \frac{dT_{\rm c}}{d\epsilon_{\rm v}}(\frac{d\epsilon_{\rm x}}{d\sigma_{\rm z}}+\frac{d\epsilon_{\rm y}}{d\sigma_{\rm z}}+\frac{d\epsilon_{\rm z}}{d\sigma_{\rm z}}) + \frac{dT_{\rm c}}{d\epsilon_{\rm tet}}(\frac{d\epsilon_{\rm z}}{d\sigma_{\rm z}}-\frac{(\frac{d\epsilon_{\rm x}}{d\sigma_{\rm z}}+\frac{d\epsilon_{\rm y}}{d\sigma_{\rm z}})}{2}),
	\label{eqtet}
\end{equation}

\begin{equation}
	\frac{dT_{\rm c}}{d\sigma_{\rm z}} = 0.00252\frac{dT_{\rm c}}{d\epsilon_{\rm v}} + 0.00558\frac{dT_{\rm c}}{d\epsilon_{\rm tet}} = 0.08755 {~\rm K/GPa}+ 0.00558\frac{dT_{\rm c}}{d\epsilon_{\rm tet}}.
	\label{Ehrenfest_eq}
\end{equation}

Thus, for the effect of the tetragonal distortion we have:	$\frac{dT_{\rm c}}{d\epsilon_{\rm tet}} = - 2.07$ K. 
Both values a very similar to the estimated one in Ref. \cite{Jerzembeck21_arxiv}. 

\subsection{Approximations for the $\langle 110 \rangle$ stress}

For the (110) stress we considered two different approximations. i) In the case of an anisotropic metal one can assume that $\sigma_{\rm xy} = \sigma_{\rm x} = \sigma_{\rm y} = \sigma_{110}$ \cite{Hicks14_Science}. ii) For the other limiting case, one assumes that the applied stress along the $\langle 110 \rangle$ directions results in stress components: $\sigma_{\rm x} = \sigma_{\rm y} = \cos(\pi/4)\sigma_{110}$, and $\sigma_{\rm xy} = \sigma_{110}$. To discriminate between these two possibilities we compared the experimental value of the Young's modulus along the $\langle 110 \rangle$ direction of $Y_{110} = 186.8$ GPa with the one obtained within these two approximations using the strain-stress matrix. For that, we need to express the change in the length of the diagonal $d$ over the changes of the $a$-axis and the angle between $d$ and $a$ (shear strain) $\pi/4-\alpha$. For zero stress: $d_0 = \sqrt{2}a_0$ and under the stress: $d'=2a'\sin{(\pi/4-\alpha)}$. For the small deformations $a' \approx  a_0(1+\epsilon_{\rm x})$ and $\sin{(\pi/4-\alpha)} \approx  (1-\epsilon_{xy})\sqrt{2}/2$. Thus, one get:
\begin{equation}
	\epsilon_{110} =  \frac{d'-d_0}{d_0} \approx \epsilon_{\rm x}-\epsilon_{xy}. 
\end{equation}
Finally, for case i) assuming compression we get: $\sigma_{110}/Y_{110}^{\rm i} = \sigma_{110}(0.01527/2-0.00307)$, hence $Y_{110}^{\rm i} \approx 219$ GPa.    
For case ii) we have: $\sigma_{110}/Y_{110}^{\rm ii} = \sigma_{110}(0.01527/2-0.00307/\sqrt{2})$, hence $Y_{110}^{\rm ii} \approx 183$ GPa.   
The second value of $Y_{110}^{\rm ii} \approx 183$ GPa is very close to the experimental $Y_{110} = 186.8$ GPa. Thus, for the further estimations, we adopted ii) as our approximation. In this case using Eq.~\ref{Eq_epsilon_xx} and ~\ref{Eq_epsilon_zz} we have:

\begin{equation}
	\frac{dT_{\rm c}}{d\sigma_{\rm 110}} = 2\frac{dT_{\rm c}}{d\epsilon_{\rm x(y)} }\frac{d\epsilon_{\rm x(y)}}{d\sigma_{\rm 110}}+\frac{dT_{\rm c}}{d\epsilon_{\rm z} }\frac{d\epsilon_{\rm z}}{d\sigma_{\rm 110}} + \frac{dT_{\rm c}}{d2\epsilon_{\rm xy} }\frac{d2\epsilon_{\rm xy}}{d\sigma_{\rm 110}}
\end{equation}

\begin{equation}
	\frac{dT_{\rm c}}{d\sigma_{\rm 110}} = 0.00307\sqrt{2}\frac{dT_{\rm c}}{d\epsilon_{\rm x(y)}}-\frac{0.00204}{\sqrt{2}}\frac{dT_{\rm c}}{d\epsilon_{\rm z}} + \frac{0.01527}{2}\frac{dT_{\rm c}}{d\epsilon_{\rm xy}} \approx 0.109 {\rm~ K/GPa} + 0.00764\frac{dT_{\rm c}}{d\epsilon_{\rm xy}} .
	\label{sigma110_der}
\end{equation}

Alternatively, the effect of the $\langle 110 \rangle$ stress can be decomposed on the fractional volume change of the unit cell $\epsilon_{\rm v}$, and the fraction of the volume-preserving tetragonal distortion $\epsilon_{\rm tet}$. Using Eq. \ref{ewV} and \ref{eqtet} we have:
\begin{equation}
	\frac{dT_{\rm c}}{d\sigma_{\rm 110}} = \frac{dT_{\rm c}}{d\epsilon_{\rm v} }\frac{d\epsilon_{\rm v}}{d\sigma_{\rm 110}}+\frac{dT_{\rm c}}{d\epsilon_{\rm tet} }\frac{d\epsilon_{\rm tet}}{d\sigma_{\rm 110}} + \frac{dT_{\rm c}}{d2\epsilon_{\rm xy} }\frac{d2\epsilon_{\rm xy}}{d\sigma_{\rm 110}} \approx 0.108 {\rm~ K/GPa} + 0.00764\frac{dT_{\rm c}}{d\epsilon_{\rm xy}},
\end{equation}
in a good agreement with Eq.~\ref{sigma110_der}.

Hence, the expected change in $T_{\rm c}$ due to shear strain is $\frac{dT_{\rm c}}{d\epsilon_{\rm xy}} = 131{\rm~GPa}~(\frac{dT_{\rm c}}{d\sigma_{110}} - 0.109 {\rm~ K/GPa}) \approx - 7.7 $ K for the measured $\frac{dT_{\rm c}}{d\sigma_{110}} = 0.05(1)$ K/GPa. 
Finally, we assumed that the contribution to $\frac{dT_{\rm TRSB}}{d\sigma_{110}}$ unrelated to the shear strain is 0.109~K/GPa (the same as for $T_{\rm c}$). Therefore, we neglect this small correction in the analysis of $\frac{dT_{\rm TRSB}}{d\sigma_{110}}$.      

\section*{ACKNOWLEDGMENTS}
The work was performed at the Swiss Muon Source (S$\mu$S), Paul Scherrer Institute (PSI, Switzerland). The work of V.G. was supported by DFG GR 4667/1. The work of A.R. was supported by the Swiss National Foundation through the Ambizione grant number No. 186043. The work is supported by JSPS KAKENHI (Nos.  JP18K04715, JP21H01033, JP22H01168, and JP22K19093), by JSPS Core-to-Core Program (No. JPJSCCA20170002), and by a JST-Mirai Program (No. JPMJMI18A3). K.I. acknowledges the support from JSPS Overseas Research Fellowships. R.S. and H-H.K acknowledges the support from DFG SFB 1143 (project ID, 247310070)) and the Würzburg-Dresden Cluster of Excellence on Complexity and Topology in Quantum Matter-ct.qmat (EXC 2147, Project ID 390858490).

\bibliography{bibliography.bib}

\end{document}